\documentclass[aps,prl,twocolumn]{revtex4-1}

\usepackage[dvips]{graphicx}
\begin{document}
\title{Exceptionally Strong Spin-Transfer in Single Ni Nanoparticles}

\author{P. Gartland}

\author{W. Jiang}

\author{D. Davidovic}
\affiliation{School of Physics, Georgia Institute of Technology, Atlanta, GA 30332}
\date{\today}

\pacs{73.23.Hk,73.63.Kv,73.50.-h}

\begin{abstract}
This letter presents  studies of spin-transfer efficiency in electron transport via discrete electron-in-a-box levels in individual nanometer-scale Ni particles at $0.06$K temperature. In a strong magnetic field, the spin-transfer rates are estimated by measuring the amplitudes of the Zeeman splitting of the levels. We find that the spin- and the charge-transfer rates are comparable, demonstrating significant enhancement of the spin-transfer efficiency compared to that in larger magnets. In a low magnetic field, we find an additional energy splitting as evidence that the spin-transfer rate is far higher than the charge-transfer rate. The effect is explained in terms of the strong mesoscopic spin-orbit torques, which are exerted on the magnetization in response to sequential electron tunneling.
\end{abstract}
\maketitle

Enhancing the efficiency of spin-transfer in electron transport through nanomagnets is of utmost importance for achieving technologically viable, nonvolatile magnetic memory and logic~\cite{katine1}. A possible solution that attracted significant interest recently is to apply spin-orbit (so) torques to change the magnetization of ferromagnets~\cite{liu,miron,miron1,pai,yamanouchi,chernyshov,cubukcu,mellnik}. Examples include the spin Hall effect in heavy metals~\cite{dyakonov,hirsch}, the Rashba effect within the free ferromagnet~\cite{edelstein,miron1,chernyshov}, and the strong so characteristics intrinsic to topological insulators~\cite{mellnik}. In this letter, we address the so-torques caused by the fluctuations of the so-fields in response to charge transport. Such fluctuations become progressively stronger in nanomagnets with significantly reduced dimensions, such as metallic ferromagnetic nanoparticles~\cite{gueron,deshmukh,jiang} and magnetic molecules~\cite{zyazin}.
In the absence of the so interaction, there will be no fundamental difference in the mechanism of spin-transfer between metallic ferromagnetic nanoparticles and larger ferromagnets. In both cases, the spin-transfer rate is on the order of $I/2eS$,~\cite{slonczewski1,berger,waintal1}, where $S$ is the total spin of the ferromagnet in units of $\hbar$, $I$ is the electron current, and $e$ is the electron charge. Any benefit derived from the reduced $S$ on the spin-transfer rate would come at the cost of a proportional decrease in the magnetization blocking temperature, weakening the required bistability of the system. Consequently, reducing the nanomagnet dimensions alone without considering other emerging effects would not be a benefit. Here we demonstrate a strong enhancement of the spin-transfer efficiency in two Ni nanoparticles in weak tunneling contact with two Al leads. In strong magnetic fields, the spin-transfer rate is smaller but already comparable to the charge-transfer rate. Even further, in weak magnetic fields, the data is consistent with a spin-transfer rate far larger than the charge-transfer rate. In the parlance of the spin Hall effect, this is analogous to a spin Hall angle much greater than one. In these nanoparticles the blocking temperature is still reduced proportionally with $S$, but the
spin-transfer rate is enhanced much faster than $1/S$. The low blocking temperature is used as a scaling-lever to provide understanding of the spin-transfer performance for higher temperature operation.

\begin{figure}[h]
    \centering
        \includegraphics[width=0.45\textwidth]{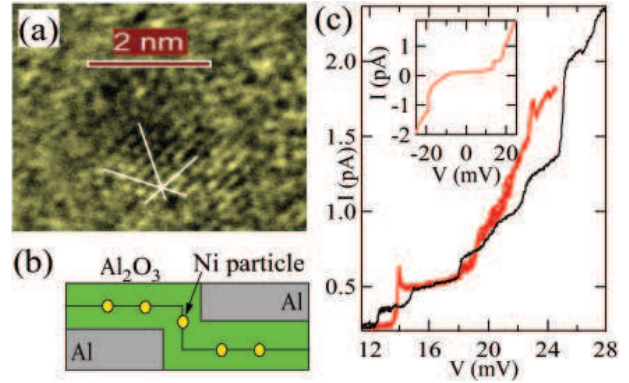}
    \label{fabandiv}
    \caption{Device fabrication geometry and current measurements. (a)Transmission Electron Microscope image of a Ni nanoparticle. (b) Double-tunneling barrier device connected to measurement apparatus through Al leads.(c) Experimental current-voltage curve for Ni sample 1. Red (black) correspond to $0$ ($11.3$) T magnetic field. Steps in the $I(V)$ curves correspond to discrete energy levels on the Ni nanoparticle. }
\end{figure}

To probe the structure of the discrete nanoparticle energy levels and their dependence on magnetic field, we perform single electron tunneling spectroscopy measurements in a dilution refrigerator. The basic device geometry is the same as that studied in previous works~\cite{wei}, and is summarized in Fig.~1. Samples are fabricated using electron-beam lithography and shadow evaporation techniques. The sample consists of two Al conducting electrodes, separated by a thin insulating layer of Al$_2$O$_3$, as shown in Fig. 1(b). Embedded within this insulating layer are Ni nanoparticles of size 1 to 4 nm. A representative Transmission Electron Microscope image of a Ni nanoparticle on an amorphous Al$_2$O$_3$ substrate is displayed in Fig. 1(a). The white lines in Fig. 1(a) indicate the primary crystal axes for the face-centered-cubic Ni.
Fig. 1(c) displays $I(V)$ curves of Ni sample 1. The red and black curves correspond to magnetic field values. The discrete steps in the $I(V)$ curves indicate the opening of another conducting channel in the nanoparticle, which arises when one of the discrete nanoparticle energy levels becomes available for tunneling. The inset of Fig. 1(c) displays the full $I(V)$ curves, which illuminates the well-known characteristic of Coulomb blockade in the low bias region.

\begin{figure}[h]
    \centering
        \includegraphics[width=0.45\textwidth]{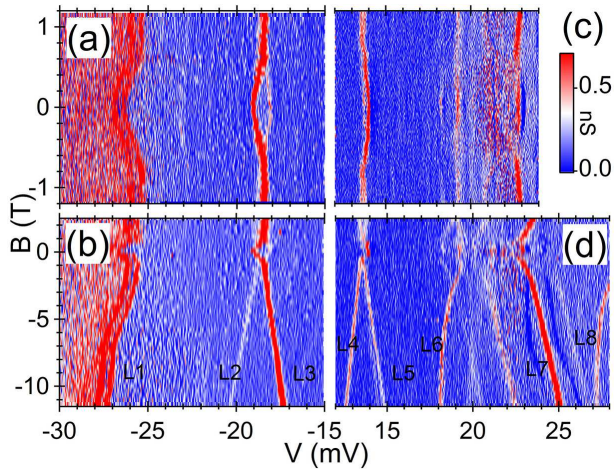}
    \label{samplespectra}
    \caption{Differential Conductance spectra of Ni sample 1. (a),(b) Spectra of Ni sample 1 in the negative bias voltage range. (c),(d) Spectra of Ni sample 1 in the positive bias range.}
\end{figure}

The energy level structure of the Ni nanoparticle is studied with differential conductance ($dI/dV$) measurements as a function of magnetic field. In these experiments, we ramp the magnetic field slowly, and sweep the bias voltage across the nanoparticle in a triangle wave fast. The current response to the applied bias voltage is measured through a current amplifier. The temperature of the mixing chamber during the field sweep is $0.06$K.
Fig. 2 and 3 displays differential conductance spectra from two exemplary Ni samples which exhibit no magnetic hysteresis. Additional data confirming the absence of magnetic hysteresis is provided in the supplementary materials.

Figs. 2(a) and 2(c) present differential conductance measurements from the low magnetic field sweeps for Ni sample 1 in the cases of negative and positive bias voltage, respectively. The energy level near -19mV splits into two branches separated by $\approx1$mV as it approaches $B=0$T. Further, the additional levels in Fig. 2(c) appear to exhibit symmetric broadening about $B=0$T, and an emerging conductance peak at +18mV near zero field. Figs. 2(b) and 2(d) contain high-magnitude field sweeps. The level that exhibited zero-field splitting in Fig. 2(a) also exhibits Zeeman splitting, as it branches into the levels L2 and L3  that vary linearly with field in the high field regime in Fig. 2(b).  For the lowest field values, the spectra have a broadened bandwidth, which focuses into more clearly-defined branches as the field is increased.
The asymmetry in the voltage values at which conduction onset occurs results from the different tunneling capacitance values in the source and drain leads. Unlike a normal metal nanoparticle,~\cite{ralph,davidovic} the Zeeman-split levels of our Ni samples do not cross at $B=0$T, but is instead offset by $\approx \pm0.5$T. The first Zeeman-split level on the negative bias side bears most of the spectral weight in the L3 branch that decreases in voltage as the field increases. The voltage-increasing branch, L2, consequently, is weak for all magnetic field values measured. The first Zeeman-split level in the positive-bias regime, however, bears comparable spectral weight in branches marked L4 and L5 for most of the field values measured. The levels at higher bias values in Fig. 2(d) exhibit both positive and negative slope, and nonlinear dependence versus magnetic field. For example, the levels L6 and L8 have stronger nonlinear dependence and negative slope versus magnetic field, while the level L7 has smaller curvature and positive slope.

\begin{figure}[h]
    \centering
        \includegraphics[width=0.45\textwidth]{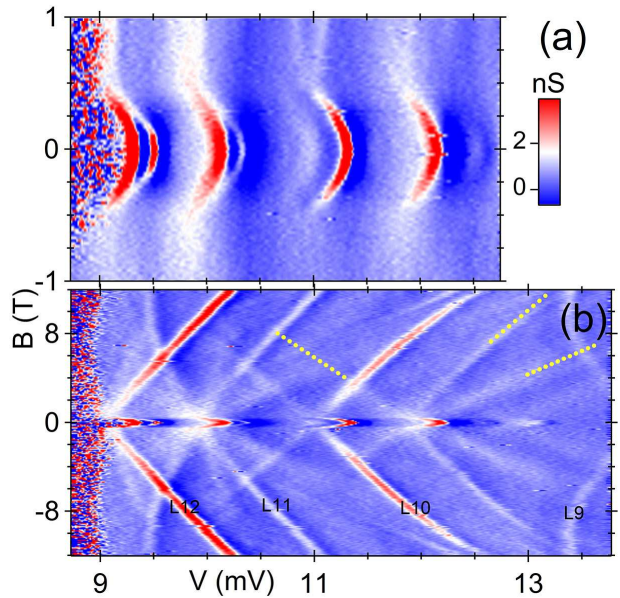}
    \label{samplespectra}
    \caption{Differential Conductance spectra of Ni sample 2. Discrete levels are indicated by L1 through L12. Dotted yellow lines indicate different Zeeman level transitions.}
\end{figure}

In Fig. 3(a) and 3(b), Ni sample 2 contains many of the same qualitative features observed in sample 1. That is, the Zeeman-split levels cross at non-zero field values.
Also, there is a single branch that bears much of the spectral weight for each Zeeman-split level.
Nonlinearities in the Zeeman-branches are apparent for higher field values.
In each the first and third levels (L12 and L10), the voltage-increasing branch carries the larger amount of spectral weight. The second and fourth levels carry smaller
amounts of spectral weight than the first and third levels, but they carry the weight more evenly among the voltage-increasing and voltage-decreasing branches.
Level L9 exhibits significant curvature in the high field, as compared with the other more linear Zeeman levels L10, L11, and L12.
By fitting a line to the Zeeman-split levels with little curvature, and correcting for capacitive division,
we estimate effective g-factors of Ni samples 1 and 2 to be 1.9 and 2, respectively.
Higher-order spin-flip transitions are also visible in Fig. 3(b), and appear as broad levels with magnitude slope $dE/dB$ corresponding to $|\Delta S_z|\approx 3/2$.
The higher order spin-flip transitions loose amplitude in magnetic field $>6$T. In Fig. 3(b), the dotted yellow lines indicate tunneling transitions corresponding to Zeeman levels with $\Delta S_z\approx\pm 1/2$, and $|\Delta S_z|\approx 3/2$.


Previous work on Co nanoparticles did not find any Zeeman splitting in discrete electron levels versus magnetic field.~\cite{gueron,deshmukh,jiang} Instead, in a strong magnetic field the discrete levels of the Co nanoparticle shifted with the same slope sign and comparable slope magnitude. At the Fermi level of a transition metal ferromagnet, the density of states of the minority electrons is much higher than that of the majority electrons. So the discrete levels shifting with the same slope sign were different minority levels of the Co nanoparticle.~\cite{gueron,deshmukh}
The absence of any measurable Zeeman splitting in the Co nanoparticle led to the conclusion that an electron with a
majority spin direction in the lead has negligibly small tunneling rate 
into the minority levels of the nanoparticle. Following that line of thought, the Zeeman levels we find in the Ni nanoparticles
demonstrate that both electrons with the minority and majority spin direction in the leads tunnel appreciably via the minority levels of the Ni-nanoparticle. The tunneling of an electron with a majority spin direction in the lead, into the minority level in the nanoparticle, involves a spin-transfer into the magnetization. So, the tunneling current via one of the two Zeeman-split levels is proportional to the spin-transfer rate, demonstrating that the spin- and the charge-transfer rates are comparable.
This conclusion is quite general in a sense that it does not rely on any particular model of the so-coupling.
 Consider for example sequential electron tunneling regime where an electron initially tunnels out of the minority level of the nanoparticle, and assume that the spin-states of the nanoparticle are close to pure spin states $|S,S_z\rangle$. In a tunneling transition between two Ni nanoparticle eigenstates, $|\alpha\rangle\to |\alpha'\rangle$, the Zeeman energy of the transition is equal to $-g\mu_BB(\langle \alpha'|S_z|\alpha'\rangle- \langle \alpha|S_z|\alpha\rangle)=-g\mu_BB\Delta \langle S_z\rangle $. The tunneling transitions that shift with $\Delta \langle S_z\rangle =\pm 1/2$, shift as the energy of the tunneling transitions $\vert S,S\rangle\to \vert S+1/2,S\pm1/2\rangle$, respectively. In the transition $\vert S,S\rangle\to \vert S+1/2,S - 1/2\rangle$, a minority electron is annihilated in the nanoparticle, while spin $-1$ is transferred into the magnetization and an electron with the majority spin direction is created in the drain lead. So, the tunneling rate associated with the transition that shifts with $\Delta \langle S_z\rangle=-1/2$ is the spin-transfer rate. In sequential electron tunneling via Zeeman split levels, the splitting of the current also depends on the resistance ratio
of the tunneling junctions,~\cite{bonet} and is generally smaller than the difference in the tunneling rates of the transitions $\vert S,S\rangle\to \vert S+1/2,S\pm 1/2\rangle$. So, in this example, the Zeeman splitting in the current sets the lower bound on the spin-transfer rate.

In the presence of so-interaction, the spin states of the nanoparticle become admixtures of pure states, which will be denoted as "$\vert S,S_z\rangle$". The tunneling transition "$\vert S,S\rangle$"$\to$"$\vert S+1/2,S-3/2\rangle$" now has nonzero probability because of the admixing. The transition energy shifts with slope corresponding to $\Delta \langle S_z\rangle=$"$\langle S+1/2,S-3/2|S_z|S+1/2,S-3/2\rangle$"$-$"$\langle S,S|S_z|S,S\rangle$", which is close to $-3/2$ if admixing is weak.
In that tunneling transition, spin "$-2$" is transferred into the magnetization from an electron tunneling out of the minority level. For the level in Fig. 3(b), the dotted yellow lines are consistent with the three tunneling transitions described by the example given in this and the previous paragraph.

To illuminate the rich structure inherent to the magnetic spectra, we use a phenomenological model based on the idea that the addition of a single electron into the nanoparticle adds a so-shift,~\cite{cehovin,usaj,brouwer3} and we assume that such a shift represents a perturbation of the nanoparticle magnetic Hamiltonian.
Due to the large mesoscopic variations in the so-shifts, it is to be expected that the particular details for each spectra may vary substantially as well. Exploring the full range of every one of these mesoscopic parameters in order to perfectly map onto each data set of experimental spectra would be computationally infeasible. Nevertheless, we can piecewise reproduce a significant number of qualitative features in Figs. 2 and 3, by choosing only one realization of the so-shift.
As the electrons tunnel sequentially via only one minority level of the nanoparticle, the nanoparticle Hamiltonian switches back and forth between $H_N$ and $H_{N+1}$, where $N$ is the number of electrons on the nanoparticle. In this work, as we have shown in previous works~\cite{gartland2}, we model $H_N$ and $H_{N+1}$ in the following forms,

$$
H_N = H(\vec{S_N})= -KS_{z,N}^2/S_N -g\mu_{B}\vec{B}\cdot\vec{S}_N
$$
$$
H_{N+1}=H(\vec{S}_{N+1})+\epsilon_{so}\left[S_{z,N+1}
+S_{x,N+1}\right]^2/2S_{N+1}^2,
$$
where $K$ is a scalar coefficient for the uniaxial anisotropy term, $g$ is the electron g-factor, $\mu_{B}$ is the Bohr magneton, $\vec{B}$ is the applied magnetic field for the Zeeman energy term, $\vec{S}_{N,N+1}$ is the vector spin operator for the nanoparticle in units of $\hbar$, and $\epsilon_{so}=-1$meV is the so-anisotropy energy added during the tunneling process. The typical value of $K\approx 10\mu$eV in the Ni nanoparticles can be obtained from the magnetic switching field.~\cite{gartland2} To obtain the tunneling spectra, we use $S=50$ and  solve the master equations numerically. In that case, the angles between the easy axes of the $N$- and $N+1$-electron particle is  $\approx\pi/6$. The intrinsic tunneling rates between the nanoparticle level and the two leads are 3MHz and 60MHz.
Fig. 4 displays the simulated differential conductance spectra.

\begin{figure}[h]
\centering
\includegraphics[width=0.45\textwidth]{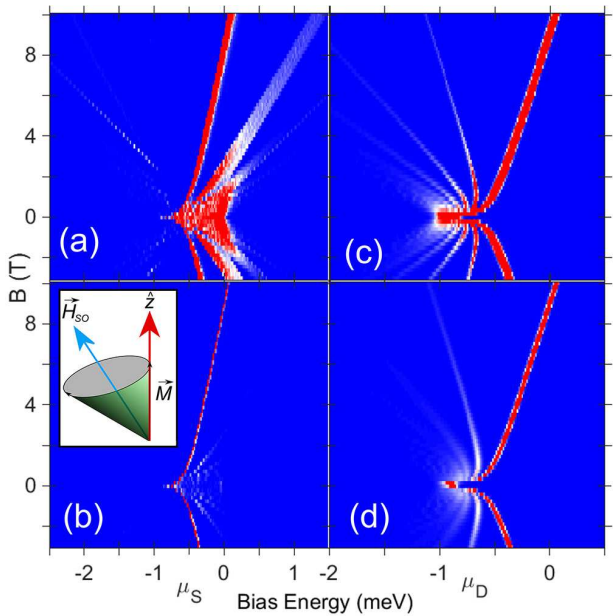}
\caption{Tunneling spectra simulations from converged Master Equations. Inset of (b) illustrates the spin-orbit torque. }
\label{fig:my_label}
\end{figure}

In Fig. 4 (a) and (b), an electron initially tunnels from the source lead into the nanoparticle, across the high and low resistance junction, respectively. The Fermi function in the drain lead, $f_d=0$, while the Fermi energy of the source lead, $\mu_s$, is swept. The Coulomb blockade region is at low values of $\mu_s$. Similarly, in Fig. 4 (c) and (d), an electron initially tunnels out of the nanoparticle into the drain lead, across the low and high resistance junction, respectively. The Fermi function in the source lead, $f_s=1$, while the Fermi level in the drain lead, $\mu_d$, is swept. The Coulomb blockade region is at high values of $\mu_d$.

In Fig. 4(a) there is energy level splitting in the vicinity of zero magnetic field, by an amount $\approx 0.6$meV. However, the zero field splitting is not apparent in Fig. 4(b)-(d). 
In Fig. 4, there are Zeeman-split branches of different spectral weights, and the
average curvature of the branches is proportional to the magnitude of $\epsilon_{so}$ for each case.
Additionally, the spectral weights redistribute amongst levels as a function of field.
That is, in Fig. 4(a), (c), and(d), multiple levels carry higher weight in the low-field, while the spectral weight redistributes toward
the energy decreasing branch in higher fields. This is reminiscent of the experimental data in
Figs. 2 and 3, where the energy levels at higher bias exhibit curvature, that is, a nonvanishing $d^2E/dB^2$, and the Zeeman levels that loose amplitude in the strong magnetic field in Fig. 3(b). The difference between the corresponding Zeeman level amplitudes in Figs. 4(a),(c) and (b),(d) explain the bias voltage asymmetry  between the corresponding Zeeman levels in Figs. 2(b) and (d). The broadening of the conductance peak about zero magnetic field in Figs. 4 (a) and (c) resemble the broadening of the conductance peak in Fig. 2(c) and 3(a).

Our phenomenological model explains the zero-field and Zeeman splitting in the measured spectra of the Ni nanoparticles, although there is no direct mapping between simulations and measurements. The value of $\epsilon_{so}$ is in reasonable agreement with theory~\cite{cehovin}. Using perturbation theory, it can be shown that the ratio of the Zeeman amplitudes in Fig. 4(a) and (c) in strong field become $(\epsilon_{so}/2\mu_B B)^2/2S$.  Reducing the value of $\epsilon_{so}$ by factor of three or more would reduce the Zeeman level amplitudes enough to be inconsistent with data. Similarly, increasing $S$ by factor of ten would produce the same effect. Our data and the model are consistent provided that the perturbation by the single-electron anisotropy is strong. In particular, $\epsilon_{so}/\sqrt{S}\gg K$. In that regime, the spin-transfer rate at low magnetic fields becomes larger than the charge transport rate.
The inset in Fig. 4(b) demonstrates this effect showing magnetization precession in the $N+1$-electron nanoparticle. If initially the magnetization is along the easy (z-) axis of the $N$-electron particle, then the single-electron anisotropy will create a so-field $H_{so}$ exerting a torque on the magnetization. In our example above, the precession of the magnetization can leave the magnetization at angle of $\approx \pi/3$ after an electron tunnels out of the nanoparticle. In that case, the spin transferred into the magnetization, in response to a single electron tunneling cycle, is $S-S\cos(\pi/3)=S/2$. We can conclude that the zero-field splitting and broadening of the measured tunneling spectra, along with the Zeeman splitting in strong magnetic field, are equivalent to  the picture of high spin-transfer efficiency in low magnetic fields, where the spin-transfer rate is far larger than the charge transport rate.


Using tunneling spectroscopy to explore the effects of the so-interaction in nanometer-scale particles offers detailed insight into the interaction of ferromagnetism, charge transport, and quantum mechanics. Due to the effect of so-fields and the torques they exert on the magnetization, the tunneling rates of spin-up and spin-down electrons become comparable in magnitude, as is manifested in the relative amplitudes in each Zeeman-split spectral branch.
The Ni nanoparticles we studied also exhibited additional spectral splitting in low magnetic field, which is attributed to the spectrum of many spin-flip excitations generated in a single electron tunneling event. While in the strong magnetic field, we find that the spin-transfer rate is smaller but comparable to the  charge-transfer rate, the zero field broadening with splitting is consistent with spin-transfer rate far higher than the charge-transfer rate. This record high spin-transfer efficiency demonstrates that it is possible, at least in principle, to substantially improve the efficiency of spin-transfer in magnetoelectronics, by taking advantage of fluctuating spin-orbit torques. The enhanced spin-transfer efficiency does not involve any heating, as is evidenced by the low electron temperature measured from the width of Zeeman levels in high magnetic field, and requires no thermal assistance. This research was supported by the US Department of
Energy, Office of Basic Energy Sciences, Division of Materials Sciences and Engineering under Award DE-FG02-06ER46281.

\bibliography{career1}

\end{document}